\documentstyle[iopfts]{ioplppt}

\begin{document}
\newcommand{\eqfeq}{$3$}
\newcommand{\eqbasis}{$6$}
\newcommand{\eqgamdec}{$7$}
\newcommand{\eqraised}{$8$}
\newcommand{\eqasLMs}{$13$}
\newcommand{\eqmetricevolution}{$19$}
\newcommand{\eqpdots}{$20$}
\newcommand{\eqconstraints}{$23$}

\jl{6}

\title{Linearization instabilities of the massive nonsymmetric 
gravitational theory}[Instabilities of massive NGT]

\author{M A Clayton\footnote{E-mail address: 
{\tt clayton@medb.physics.utoronto.ca}}}
\address{Department of Physics, University of Toronto, Toronto, ON, 
Canada M5S 1A7}

\begin{abstract}
The massive nonsymmetric gravitational theory is shown to possess a 
linearization instability at purely GR field configurations, disallowing 
the use of the linear approximation in these situations.
It is also shown that arbitrarily small antisymmetric sector Cauchy data 
leads to singular evolution unless an {\it ad hoc} condition is imposed 
on the initial data hypersurface.
\end{abstract}

\pacs{04.50.+h, 04.25.Nx, 04.20.Ex, 11.10.Jj}


\section{Introduction\label{sect:intro}}

The Nonsymmetric Gravitational Theory (NGT) (Moffat 1979, 1984, 1991) 
grew out of a reinterpretation of Einstein's unified field theory 
(Einstein 1945; Einstein and Straus 1946) as a theory of gravitation 
alone, thereby bypassing the problems associated with the interpretation 
of the additional structure as representing the electromagnetic field 
(Infeld 1950, Callaway 1953).
There were indications of problems even with this reinterpretation 
(Kunstatter \etal 1984; Kelly 1991, 1992) which were clarified by Damour, 
Deser and McCarthy (1992, 1993) who showed that the wave solutions of the 
weak field equations did not decrease at large distances from the source 
along the forward light cone (discussed in more detail in Clayton (1996a)).

The theory was subsequently altered into what will be referred to as the 
massive Nonsymmetric Gravitational Theory (mNGT) (L\'{e}gar\'{e} and Moffat 
1995; Moffat 1995a, b) by requiring that the linearized field 
equations reduce to those of a massive Kalb-Ramond (1974) field, guaranteeing 
that the linearized fields are well-behaved asymptotically far from the 
source (Clayton 1996a).
The action for the Massive theory is ($16\pi G=c=1$)
\begin{eqnarray}
\label{eq:NGTAction}
S_{\rm ngt}=\int\d^4x\,E\bigl[
-{\bf g}^{AB}R_{AB}^{\rm ns}
-{\bf g}^{AB}\nabla_{e_{[A}}[W]_{_{B]}}
+{\bf l}^A\Lambda_A\nonumber \\
\qquad\qquad\qquad{}+\case{3}{8}{\bf g}^{(AB)}W_A W_B
+\case{1}{4}m^2{\bf g}^{[AB]}{\rm g}_{[AB]}\bigr],
\end{eqnarray}
where the first three terms are identical to the vacuum UFT action, the last 
term is the Bonnor (1954) term, and the remaining contribution is the new 
feature of mNGT designed to obtain the correct linearisation.
(Note that the Lagrange multiplier $l^A$ is {\it not} a new feature in the
NGT action; if one uses the connection as defined by  Moffat (1991), it 
would not appear.)

The convention of this work is to define the inverse of the (in general 
nonsymmetric: ${\rm g}_{AB}\neq{\rm g}_{BA}$) fundamental tensor by 
${\rm g}^{AB}{\rm g}_{BC}={\rm g}_{CB}{\rm g}^{BA}=\delta^A_C$.
Densities with respect to the determinant of the fundamental tensor 
${\rm g}:=\det[{\rm g}_{AB}]$ will be indicated by boldface: 
${\bf T}^{AB}:=\sqrt{-{\rm g}}{\rm T}^{AB}$, and the symmetric and 
antisymmetric part of a tensor will be indicated by 
$T_{(AB)}:=(T_{AB}+T_{BA})/2$ and $T_{[AB]}:=(T_{AB}-T_{BA})/2$ respectively.
A torsion-free covariant derivative is employed, characterised by the 
connection coefficients that define the parallel transport of the frame by 
$\nabla_{e_A}[e]_B=\Gamma^C_{AB}e_C$ (which is not required 
to be compatible with any tensor), and what are normally 
treated as the antisymmetric components of the connection coefficients 
are considered as a separate tensor $\Lambda^A_{BC}$, the trace of which is
defined by $\Lambda_A:=\Lambda^B_{AB}$.
The action (\ref{eq:NGTAction}) has been written in a general linear frame, 
where the basis vectors are related to a coordinate basis via a vierbein 
$e_A={E_A}^{\mu}\partial_\mu$, generally leading to nontrivial structure 
constants $[e_A,e_B]={C_{AB}}^Ce_C$.

The NGT Ricci tensor that appears in the action can be split up into 
two contributions: $R_{AB}^{\rm ns}=R_{AB}+R^\Lambda_{AB}$; the 
first is identified as the Ricci tensor ({i.e.}, it reduces to 
the GR Ricci tensor in the limit of vanishing antisymmetric sector), 
and the second contains contributions from the antisymmetric tensor 
field $\Lambda^A_{BC}$
\numparts
\begin{eqnarray}
\label{eq:ncbRicci}
\eqalign{
R_{AB}=e_C[\Gamma^C_{BA}]
-e_B[\Gamma^C_{CA}]
-\case{1}{2}e_A[\Gamma^C_{BC}]
+\case{1}{2}e_B[\Gamma^C_{AC}]\\
\qquad\qquad{ }+\Gamma^E_{BA}\Gamma^C_{CE}
-\Gamma^E_{CB}\Gamma^C_{EA}
+\Gamma^E_{[AB]}\Gamma^C_{EC},}\\
\label{ncbRNgt}
R^\Lambda_{AB}=
\nabla_{e_C}[\Lambda]^C_{AB}
+\nabla_{e_{[A}}[\Lambda]_{_{B]}}
+\Lambda^C_{AD}\Lambda^D_{BC}.
\end{eqnarray}
\endnumparts
In addition to the vanishing of the torsion tensor
\numparts
\begin{equation}\label{torsion}
T^A_{BC}:=\Gamma^A_{BC}-\Gamma^A_{CB}-{C_{BC}}^A=0,
\end{equation}
the field equations of mNGT take on the form (Clayton 1996a):
\begin{eqnarray}
\label{mNGTF:b}
\Lambda_A=0, \\
\label{mNGTF:d}
{\bf l}^A=\case{1}{4}{\bf g}^{(AB)}W_B,\\
\label{mNGTF:c}
\nabla_{e_B}[{\bf g}]^{[AB]}=\case{3}{4}{\bf g}^{(AB)}W_B,\\
\label{eq:compat}
\nabla_{e_C}[{\bf g}]^{AB}
-{\bf g}^{DB}\Lambda^A_{CD}
-{\bf g}^{AD}\Lambda^B_{DC}
+\case{1}{2}\delta^{[A}_C\delta^{B]}_D{\bf g}^{(DE)}W_E=0,\\
\label{mNGTF:a}
R^{\rm ns}_{AB}+\nabla_{e_{[A}}[W]_{_{B]}}
-\case{3}{8}W_{A}W_{B}
-\case{1}{4}m^2M_{AB}=0, 
\end{eqnarray}
\endnumparts
where the mass tensor appearing in the last of these is defined by
\begin{equation}\label{equation:mass tensor}
M _{AB}={\rm g}_{[AB]}-{\rm g}_{CA}{\rm g}_{BD}{\rm g}^{[CD]}
+\case{1}{2}{\rm g}_{BA}{\rm g}^{[CD]}{\rm g}_{[CD]}.
\end{equation}

By requiring that the linearisation of the antisymmetric sector take on the
form of a Kalb-Ramond field, the field equations have picked up three 
additional constraints at the linearized level, which have since been shown 
to be absent in the nonperturbative theory (Clayton 1995, 1996b).
The purpose of the present work is to examine the form of the exact 
field equations (\eqfeq) near initially GR field configurations, demonstrating 
the fact that the exact evolution of the system is not well approximated 
by the linearized system.
It will therefore have been shown that mNGT is not linearization stable 
about GR spacetimes, and so the linearization which guided the 
construction of the action does not describe the theory even in an 
approximate sense.
Furthermore, it will be shown that unless an {\em ad hoc} constraint is 
imposed on the initial data hypersurface, unboundedly large velocities 
occur for arbitrarily small antisymmetric sector fields.
This indicates a Cauchy instability (as defined in Hawking and Ellis 
(1973)), since the field configuration at an infinitesimally small time 
later will not depend smoothly on the initial data near a GR 
configuration.

The paper will consist of two main sections: the first will review the field 
equations of mNGT in $3+1$ decomposed form, and the second will explore 
the above issues, making it clear how and why the naive linearization 
fails.
The discussion of the surface decomposition of mNGT is fairly terse, and 
further details about the formalism are given in (Clayton 1995, 1996b).
This construction is a straightforward extension of that given by 
Isenberg and Nestor (1980), although it must be admitted that the structure
of NGT introduces a great deal of algebraic complication that necessitates
the introduction of many auxiliary fields in order to present results
in a reasonably compact form.
In addition, in order to work consistently with a surface adapted basis
in NGT, it was necessary to develop a `moving frame' formalism for NGT
(Clayton 1996a).

\section{The Surface-Decomposed Field Equations\label{sect:FEq}}

What will be given here is a summary of the space and time decomposition 
that was used to consider the Cauchy problem in NGT (Clayton 1995, 1996b).
Spacetime has been foliated into spacelike hypersurfaces 
$({\bf M},{\rm g})\sim\{(\Sigma_t,{\rm g}_t)\vert t\in 
I\subset \Bbb{R}\}$, where the metric of spacetime has been 
operationally identified with the symmetric components of the inverse of 
the fundamental tensor of NGT ${\rm g}^{(AB)}$. 
Configuration space has therefore been chosen to consist of 
${\rm g}^{[\perp a]}$ and ${\rm g}^{ab}$, where 
$\{a,b,\ldots\}\in\{1,2,3\}$ indicate components of a tensor on $\Sigma$ 
expanded in a coordinate basis, and the index `$\perp$' indicates a 
component normal to the hypersurface as defined by ${\rm g}^{(AB)}$.

This surface compatible basis is designed so that the symmetric components 
of the inverse of the fundamental tensor take on the ADM form (Arnowitt 
\etal 1959): 
${\rm g}^{-1}_{{}_{(\;)}}=e_\perp\otimes e_\perp-\gamma^{(ab)}e_a\otimes 
e_b$, where the basis may be written in terms of the coordinate basis 
($\partial_{{t}},\partial_a$) as: 
$e_\perp=\frac{1}{N}\partial_{{t}}-\frac{N^a}{N}\partial_a$, 
$e_a=\partial_a$, and the covector bases as: $\theta^\perp=Nd{t}$, 
$\theta^a=dx^a+N^ad{t}$.
The time vector is decomposed as ${\sf t}=Ne_\perp+\vec{N}$, where 
$(N,N^a)$ are the lapse function and shift vector respectively, and the 
notation $\vec{N}:=N^ae_a$ has been employed.
The non-vanishing structure constants are easily found to be:
\begin{equation}\label{Cs}
{C_{\!\perp a}}^\perp=e_a[\ln(N)],\qquad
{C_{\!\perp a}}^b=\frac{1}{N}e_a[N^b],
\end{equation}
and the fundamental tensor and its inverse become:
\numparts
\begin{eqnarray}\label{eq:fmet}
\eqalign{
\fl
{\rm g}=F\theta^\perp\otimes\theta^\perp
-(\alpha_a-\beta_a)\theta^\perp\otimes\theta^a
-(\alpha_a+\beta_a)\theta^a\otimes\theta^\perp
-G_{ab}\theta^a\otimes\theta^b, \\
\fl
{\rm g}^{-1}=e_\perp\otimes e_\perp
+B^ae_\perp\otimes e_a-B^ae_a\otimes e_\perp
-\gamma^{ab}e_a\otimes e_b,}
\end{eqnarray}
where:
\begin{equation}\label{invmetcomp}
F:=1+G_{(ab)}B^aB^b,\qquad 
\alpha_a:=G_{[ab]}B^b,\qquad 
\beta_a:= G_{(ab)}B^b,
\end{equation}
from which one finds that $\alpha_aB^a=0$.
The spatial part of the fundamental tensor is given by
\begin{equation}\label{eq:ginvG}
(\gamma^{ac}-B^a B^c)G_{cb}=
G_{bc}(\gamma^{ca}-B^c B^a)=
\delta^a_b.
\end{equation}
\endnumparts
Note that invertibility of the fundamental tensor in (\ref{eq:fmet}) is 
required in order that the volume element in the action be nondegenerate, 
which in turn requires that $G_{ab}$ exist.

The spacetime covariant derivatives have been decomposed as usual 
(Isenberg and Nester 1980) into surface covariant derivatives (written 
as $\nabla^{(3)}_a$) whose action on basis vectors in $T\Sigma$ (the 
space of vectors tangent to $\Sigma$) is given by 
$\nabla^{(3)}_a[e]_b=\Gamma^c_{ab}e_c$, and contributions from 
derivatives off of $\Sigma$ which will be written in terms of the 
surface tensors:
\numparts
\begin{eqnarray}
\label{eq:gam dec}
\eqalign{
\Gamma:=\Gamma^\perp_{\perp\perp},\qquad
c_a:=\Gamma^\perp_{a\perp},\qquad
a_a:=\Gamma^\perp_{\perp a},\qquad
\sigma^a:=\Gamma^a_{\perp\perp},\\
w^a_{\;b}:=\Gamma^a_{\perp b},\qquad
u^a_{\;b}:=\Gamma^a_{b\perp},\qquad
k_{ab}:=\Gamma^\perp_{ab}.}
\end{eqnarray}
In GR, (\ref{eq:gam dec}) are related by the algebraic compatibility 
conditions: $\sigma^a=\gamma^{ab}a_b$, $c_a=0$, $\Gamma=0$, and 
$u^a_{\;b}=\gamma^{ac}k_{bc}$, as well as the conditions that enforce 
the vanishing of torsion: $a_a={C_{\perp a}}^\perp$, 
$w^a_{\;b}=u^a_{\;b}+{C_{\perp b}}^a$, $\Gamma^a_{[bc]}=0$, and 
$k_{[ab]}=0$.
The antisymmetric tensor $\Lambda^A_{BC}$ is decomposed as:
\begin{equation}\label{lam dec}
b_a:=\Lambda^\perp_{a\perp},\qquad
j_{ab}:=\Lambda^\perp_{ab},\qquad
v^a_{\;b}:=\Lambda^a_{\perp b},\qquad
\lambda^a_{bc}:=\Lambda^a_{bc},
\end{equation}
\endnumparts
where by definition $j_{ab}$ is an antisymmetric surface tensor.
The vector field $W_A$ that appears in (\ref{eq:NGTAction}) is decomposed 
as $W_A=(W,W_a)$, and the traceless part of $\lambda^a_{bc}$ is defined by $\lambda^a_{Tbc}=\lambda^a_{bc}-\delta^a_{[b}b_{c]}$.
It is also useful to make the following definitions:
\numparts
\begin{eqnarray}
K_{ab}:=k_{ab}+j_{ab},\\
\eqalign{
k^a_{\;b}:=\case{1}{2}(\gamma^{ac}K_{bc}+\gamma^{ca}K_{cb})
=\gamma^{(ac)}k_{bc}+\gamma^{[ac]}j_{bc},\\
j^a_{\;b}:=\case{1}{2}(\gamma^{ac}K_{bc}-\gamma^{ca}K_{cb})
=\gamma^{[ac]}k_{bc}+\gamma^{(ac)}j_{bc},\\
}
\end{eqnarray}
\endnumparts
the first of which is defined for notational convenience, and the 
remaining are symmetric and antisymmetric sector contributions from 
tensors that have had indices `raised' by $\gamma^{ab}$.

In the Hamiltonian formalism of mNGT (Clayton 1995, 1996b) the 
canonical momenta are weakly equivalent to the densitized components of 
the fundamental tensor appearing in (\ref{eq:fmet}), and the 
set of conjugate pairs of phase space coordinates are
\begin{equation}\label{eq:Can coord}
\{({\bf B}^a,\overline{W}_a),(\bgamma^{ab},K_{ab})\},
\end{equation}
where the field
\begin{equation}\label{eq:Wbar}
\overline{W}_a:=-W_a+2b_a,
\end{equation}
naturally appears as as the momentum conjugate to ${\bf B}^a$.
(In (Clayton 1995, 1996b) there is an additional pair of symmetric 
sector coordinates that does not exist in typical Cauchy analyses of GR 
which is removable by making imposing the second class 
constraints: $p\approx\sqrt{-{\rm g}}[{\bf B}^a,\bgamma^{ab}]$, and 
$u\approx \gamma^{ab}K_{ab}$ strongly.)

The fact that the spacetime covariant derivative has been defined to be 
torsion-free (\ref{torsion}) as in GR results in 
$k_{[ab]}=\Gamma^c_{[ab]}=0$, now combined with:
\begin{equation}\label{eq:torsion free}
a_a\approx c_a+{C_{\perp a}}^\perp,\qquad
w^a_{\;b}:= u^a_{\;b}+{C_{\perp b}}^a,
\end{equation}
and equation (\ref{mNGTF:b}) gives:
\begin{equation}\label{eq:lam A}
v^a_{\;a}\approx 0,\qquad b_a\approx -\lambda^b_{ab}.
\end{equation}

The symmetric sector Lagrange multipliers that appear 
in (\eqgamdec) may be solved for in terms of the antisymmetric 
sector Lagrange multipliers and Cauchy data as:
\numparts
\begin{eqnarray}
%
\eqalign{
\bGamma
\approx {\bf B}^ab_a
-\case{1}{8}\bgamma^{(ab)}\beta_a(\overline{W}_b-2b_b),\\
%
\bsigma^a
\approx \bgamma^{(ab)}a_b
-{\bf B}^bv^a_{\;b}-\bgamma^{[ab]}b_b,  \\
%
{\bf c}_a
\approx -j_{ab}{\bf B}^b
+\case{1}{8}\beta_a{\bf W}
+\case{1}{8}G_{[ab]}\bgamma^{(bc)}(\overline{W}_c-2b_c), \\
%
{\bf u}^a_{\;b}
\approx {\bf B}^ab_b
+{\bf B}^c\lambda^a_{cb}
+{\bf k}^a_{\;b},}
\end{eqnarray}
and using the solution for $v^a_{\;b}$ given in (\ref{eq:v rel}) one finds 
the surface compatibility condition that determines $\Gamma^a_{bc}$ in terms
of Cauchy data:
\begin{equation}\label{eq:Gamma rel}
\nabla^{(3)}_{c}[\bgamma]^{(ab)}\approx
\bgamma^{(ab)}c_c
+2{\bf B}^{(a}v^{b)}_{\;c}
+\bgamma^{[ad]}\lambda^b_{dc}
+\bgamma^{[db]}\lambda^a_{cd}.
\end{equation} 
(The Lagrange multipliers $v^a_{\;b}$ must be replaced in 
(\ref{eq:Gamma rel}) in order to solve for the surface connection 
coefficients, since $v^a_{\;b}$ in (\ref{eq:v rel}) has explicit dependence 
on $\Gamma^a_{bc}$.) 
In the antisymmetric sector one can solve for:
\begin{equation}\label{eq:v rel}
%
{\bf W}\approx\case{4}{3}\nabla^{(3)}_a[{\bf B}]^a,\qquad
%
{\bf v}^a_{\;b}
\approx \nabla^{(3)}_{b}[{\bf B}]^a
+{\bf j}^a_{\;b}
-\case{1}{4}\delta^a_b{\bf W}, 
\end{equation}
leaving
\begin{equation}\label{eq:lambda rel}
\fl
\nabla^{(3)}_{c}[\bgamma]^{[ab]}\approx
-2{\bf B}^{[a}u^{b]}_{\;c}
+\bgamma^{[ab]}c_c
+\bgamma^{(ad)}\lambda^b_{dc}
+\bgamma^{(db)}\lambda^a_{cd}
+\case{1}{2}\delta^{[a}_c\bgamma^{(b]e)}(\overline{W}_e-2b_e)
\end{equation}
\endnumparts
to determine the Lagrange multipliers $b_a$ and $\Lambda^a_{Tbc}$.

The general solution of the full set of Lagrange multipliers is highly 
nontrivial, but will not be required here.
Only the relation that determines $b_a$ in terms of Cauchy data 
(isolated by taking the trace of (\ref{eq:lambda rel}) and using the 
remainder of (\eqasLMs) in order to replace the Lagrange 
multipliers with Cauchy data),
\begin{eqnarray}\label{eq:full b}
\eqalign{
\fl
\sqrt{-{\rm g}}[B^aB^b-\case{1}{4}\gamma^{[ac]}G_{[cd]}\gamma^{(db)}]
b_b\approx 
\case{1}{2}\bgamma^{(ab)}\overline{W}_b
-\case{1}{8}\bgamma^{[ab]}G_{[bc]}\gamma^{(cd)}\overline{W}_d\\
{}+e_b[\bgamma^{[ab]}]
+k^b_{\;b}{\bf B}^a-k^a_{\;b}{\bf B}^b
+\bgamma^{[ab]}j_{bc}B^c
-\case{1}{6}\gamma^{[ab]}\beta_b\nabla^{(3)}_c[{\bf B}]^c,}
\end{eqnarray}
will be necessary in order to demonstrate the result.
It is this rather surprising ability to isolate a relation that 
determines $b_a$ in terms of Cauchy data that allows the analysis in the 
next section to be performed in a fairly straightforward manner.

It will be useful to define the two tensors: 
\numparts
\begin{equation}\label{eq:Os}
O_1^{ab}:=\gamma^{(ab)}-B^aB^b,\qquad
O_2^{ab}:=B^aB^b-\case{1}{4}\gamma^{[ac]}G_{[cd]}\gamma^{(db)},
\end{equation}
the inverses of which will be denoted $O^{-1}_{1ab}$ and $O^{-1}_{2ab}$ 
respectively, given in general by:
\begin{eqnarray}\label{eq:left inv}
\eqalign{
O^{-1}_{1ab}=S_{ab}+\frac{B_a B_b}{1-B\cdot B}, \\
O^{-1}_{2ab}=\frac{1}{\gamma\cdot G}\Bigl[
S_{ab}-\frac{B_aG_b+G_aB_b}{G\cdot B}
+\frac{B\cdot B+\gamma\cdot G}{(\gamma\cdot B)(G\cdot B)}
G_a\gamma_b\Bigr].}
\end{eqnarray}
\endnumparts
Indices have been `raised' or `lowered' by $\gamma^{(ab)}$ and its 
inverse $S_{ab}$ (for example, $G_a:=S_{ab}G^b$, and 
$\gamma^{(ab)}S_{bc}=\delta^a_c$), and the notation 
$B\cdot G:=B^aG_a$ has been employed.
The $3$-dimensional antisymmetric symbol $\epsilon_{abc}$ and the 
surface volume element $\sqrt{S}:=\sqrt{\det S_{ab}}$ have been used 
to define:
\begin{eqnarray}\label{eq:exact defns}
\eqalign{
\gamma^{[ab]}=\frac{2}{\sqrt{S}}\epsilon^{abc}\gamma_c,\qquad 
\gamma_a:=\frac{1}{4}\sqrt{S}\epsilon_{abc}\gamma^{[bc]},\\
G_{[ab]}=2\sqrt{S}\epsilon_{abc}G^c,\qquad 
G^a:=\frac{1}{4\sqrt{S}}\epsilon^{abc}G_{[bc]},}
\end{eqnarray}
where $\epsilon_{123}=\epsilon^{123}=+1$.
Clearly in the limit of vanishing antisymmetric sector, $O^{-1}_{2ab}$ 
in (\ref{eq:left inv}) is not in general well-behaved due to the 
presence of $(\gamma\cdot G)^{-1}$ which is singular in 
that limit.

Using these and the definition
\begin{equation}\label{eq:Xi def}
\fl
\case{1}{2}\sqrt{-{\rm g}}\Xi^a:=
e_b[\bgamma^{[ab]}]+k^b_{\;b}{\bf B}^a-k^a_{\;b}{\bf B}^b
+\bgamma^{[ab]}j_{bc}B^c
-\case{1}{6}\gamma^{[ab]}\beta_b\nabla^{(3)}_c[{\bf B}]^c,
\end{equation}
(\ref{eq:full b}) is written as
\numparts
\begin{equation}\label{eq:full b2}
O_2^{ab}b_b\approx 
\case{1}{2}O^{ab}_1\overline{W}_b+\case{1}{2}O^{ab}_2\overline{W}_b
+\case{1}{2}\Xi^a,
\end{equation}
the solution of which is (in terms of the phase space coordinates)
\begin{equation}\label{eq:b sol}
b_a\approx \case{1}{2}\overline{W}_a 
+\case{1}{2}O^{-1}_{2ab}O^{bc}_1\overline{W}_c
+\case{1}{2}O^{-1}_{2ab}\Xi^b.
\end{equation}
In terms of the configuration space variables, equation (\ref{eq:full b}) 
becomes
\begin{equation}\label{eq:Con b}
O^{ab}_1b_b
=\case{1}{2}O^{ab}_1W_b+\case{1}{2}O^{ab}_2W_b+\case{1}{2}\Xi^a,
\end{equation}
where $W_a$ must be solved for in terms of time derivatives of the 
configuration space variables $({\bf B}^a,\bgamma^{ab})$ in 
(\ref{eq:metric evolution:b}) below.
The solution of (\ref{eq:Con b}) may then be determined to be
\begin{equation}\label{eq:b solution}
b_a=\case{1}{2}W_a+\case{1}{2}O^{-1}_{1ab}O_2^{bc}W_c
+\case{1}{2}O^{-1}_{1ab}\Xi^b,
\end{equation} 
resulting in
\begin{equation}\label{eq:Wbar sol}
\overline{W}_a=-W_a+2b_a=O^{-1}_{1ab}O_2^{bc}W_c+O^{-1}_{1ab}\Xi^b.
\end{equation}
\endnumparts

The evolution of the canonical momenta as determined from Hamilton's 
equations or, equivalently, from the compatibility conditions in 
(\ref{eq:compat}) that involve time derivatives of the fundamental tensor, 
is given by:
\numparts
\begin{eqnarray}
\fl
\label{eq:metric evolution:b}
\dot{{\bf B}}^a\approx \pounds^{(3)}_{\vec{N}}[{\bf B}]^a
-N(\nabla^{(3)}_{b}[\bgamma]^{[ab]}
+\bgamma^{[ab]}{C_{\!\perp b}}^\perp
+\case{3}{4}\bgamma^{(ab)}(\overline{W}_b-2b_b)),\\
\fl
\label{eq:metric evolution:c}
\dot{\bgamma}^{(ab)}\approx \pounds^{(3)}_{\vec{N}}[\bgamma]^{(ab)}
+N(\bgamma^{(ab)}(\Gamma+u)
-2(\bgamma^{(c(a)}u^{b)}_{\;c}
-\bgamma^{[c(a]}v^{b)}_{\;c}),\\
\fl
\label{eq:metric evolution:d}
\dot{\bgamma}^{[ab]}\approx \pounds^{(3)}_{\vec{N}}[\bgamma]^{[ab]}
-N({\bf B}^a\sigma^b-{\bf B}^b\sigma^a-\bgamma^{[ab]}(\Gamma+u)
-2(\bgamma^{[c[a]}u^{b]}_{\;c}
-\bgamma^{(c[a)}v^{b]}_{\;c}),
\end{eqnarray}
\endnumparts
where $\pounds^{(3)}$ is the surface Lie derivative.
Partial derivatives with respect to the time coordinate have been 
represented by an overdot (e.g. $\dot{{\bf B}}^a$), and are 
equivalent to the properly defined Lie derivative off of the surface 
(Isenberg and Nester 1980; Clayton 1995, 1996b).
The evolution of the coordinates (from Hamilton's equations or 
(\ref{mNGTF:a})) is determined to be:
\numparts
\begin{eqnarray}
\label{eq:p dots:b}
\dot{\overline{W}}_a\approx \pounds^{(3)}_{\vec{N}}[\overline{W}]_a
-2N{\cal Z}_{[a\perp]}
+\case{1}{2}m^2NM _{[a\perp]}, \\
\label{eq:p dots:c}
\dot{K}_{ab}\approx \pounds^{(3)}_{\vec{N}}[K]_{ab}-N{\cal Z}_{ab}
+\case{1}{4}m^2NM _{ab},
\end{eqnarray}
\endnumparts
where ${\cal Z}_{AB}$ as determined from (\ref{mNGTF:a}) are given by:
\numparts\label{eq:Z}
\begin{eqnarray}
\label{eq:Z ab symm}
{\cal Z}_{(ab)}:=
R^{(3)}_{(ab)}-\nabla^{(3)}_{{(a}}[a]_{b)}+(\Gamma+u)k_{ab}
-a_aa_b-k_{cb}u^c_{\;a}-k_{ac}u^c_{\;b} \nonumber \\
\qquad{}+b_ab_b
-j_{ac}v^c_{\;b}-j_{bc}v^c_{\;a}+\lambda^c_{ad}\lambda^d_{bc}
-\case{3}{8}(\overline{W}_a-2b_a)(\overline{W}_b-2b_b),\\
\label{eq:Za aperp}
{\cal Z}_{[a\perp]}:=
\case{1}{2}\nabla^{(3)}_{a}[W]+\case{1}{2}W(a_a-c_a)
-\nabla^{(3)}_{b}[v]^b_{\;a}\nonumber \\
\qquad{}-j_{ab}\sigma^b +(c_b-a_b)v^b_{\;a}
+ub_a-u^b_{\;a}b_b-u^b_{\;c}\lambda^c_{ab},\\
{\cal Z}_{[ab]}:=
-\nabla^{(3)}_{[a}[\overline{W}]_{b]}+2\nabla^{(3)}_{[a}[b]_{b]}
+\nabla^{(3)}_{c}[\lambda]^c_{ab}
+(\Gamma+u)j_{ab}+a_c\lambda^c_{ab}\nonumber \\
\qquad{}+a_ab_b-a_bb_a-j_{cb}u^c_{\;a}-j_{ac}u^c_{\;b}
-k_{ca}v^c_{\;b}+k_{cb}v^c_{\;a}.
\end{eqnarray}
\endnumparts
In (\ref{eq:Z ab symm}) the surface Ricci tensor is defined from 
(\ref{eq:ncbRicci}) to be
\begin{eqnarray}\label{eq:Sigma Ricci}
R^{(3)}_{ab}&=e_c[\Gamma^c_{ba}]-\case{1}{2}
(e_b[\Gamma^c_{ca}]
+e_a[\Gamma^c_{bc}])
+\Gamma^c_{ba}\Gamma^d_{dc}
-\Gamma^c_{da}\Gamma^d_{bc},
\end{eqnarray}
which is constructed from contractions of the intrinsic curvature of 
$\nabla^{(3)}$ on $\Sigma$.

The remaining field equations in (\ref{mNGTF:a}) are the Gauss relation or 
Hamiltonian constraint (Clayton 1995, 1996b)
\numparts
\begin{eqnarray}
\label{eq:SH constraint}
\eqalign{
\fl
{\cal H}\approx -(u{\bf u}-u^a_{\;b}{\bf u}^b_{\;a}
+v^a_{\;b}{\bf v}^b_{\;a}+\bgamma^{(ab)}R^{(3)}_{(ab)})
+\case{1}{4}m^2(pM_{\perp\perp}+\bgamma^{ab}M_{ab})\\
{}-\bgamma^{(ab)}(\lambda^c_{Tad}\lambda^d_{Tbc}
-\case{3}{8}\overline{W}_a\overline{W}_b
+\case{3}{2}\overline{W}_ab_b) 
+\case{3}{8}\sqrt{-{\rm g}}(W)^2 \\
{}-\bgamma^{[ab]}(-\nabla^{(3)}_a[\overline{W}]_b+2\nabla^{(3)}_a[b]_b
+\nabla^{(3)}_c[\lambda]^c_{ab}) 
-2u{\bf B}^ab_a \\
{}+\nabla^{(3)}_a[{\bf B}^bv^a_{\;b}+\bgamma^{[ab]}b_b]
-c_a({\bf B}^bv^a_{\;b}+\bgamma^{[ab]}b_b)
\approx 0,}
\end{eqnarray}
and the Codacci relations or momentum constraints
\begin{eqnarray}\label{eq:SM constraint}
\fl
{\cal H}_a
\approx{\bf B}^{b}e_{a}[\overline{W}_{b}]
-e_{b}[{\bf B}^{b}\overline{W}_{a}]
+\bgamma^{bc}e_{a}[K_{bc}]+\sqrt{-{\rm g}}e_a[u]
-2e_{b}[{\bf k}^b_{\;a}]\approx 0.
\end{eqnarray}
\endnumparts
Diffeomorphism invariance of the mNGT action guarantees that the Poisson 
brackets of the Hamiltonian and momentum constraints satisfy the same 
closing relations as GR (Teitelboim 1973, 1980; Isenberg and Nester 
1980), where the fact that $\gamma^{(ab)}$ appears in the Poisson 
bracket of the Hamiltonian constraint is a consequence of the choice of 
using ${\rm g}^{(AB)}$ to define the unit normal to $\Sigma$, and is 
discussed in more detail in (Clayton 1996b).

One now has the exact evolution equations of mNGT, written in surface 
decomposed form that is suitable for considering the Cauchy problem in 
more detail.
As a first-order system, one would first solve (\eqasLMs) for 
the Lagrange multipliers replacing them everywhere in Hamilton's 
equations (\eqmetricevolution) and (\eqpdots) and 
the constraints (\eqconstraints).
There would then be four undetermined Lagrange multipliers ($N$ and 
$N^a$) and four diffeomorphism constraints (\eqconstraints) on the Cauchy 
data (\ref{eq:Can coord}). 
To consider the system as second-order, one would go further and solve 
(\eqmetricevolution) for the canonical coordinates 
$(\overline{W}_a, K_{ab})$, replacing them in the constraints 
(\eqconstraints)as well as (\eqpdots), the collection of which would then 
become the second-order Euler-Lagrange equations of mNGT. 

\section{Linearization Instability\label{sect:E-L}}

Considering the complexity of the field equations, it would be somewhat 
too optimistic to attempt to draw definite conclusions about global 
solutions of the mNGT field equations (see Moffat (1995c) for some results).
Instead it will be sufficient to consider the instantaneous problem 
here, considering Cauchy data for which the antisymmetric sector will be 
chosen to be an arbitrarily small perturbation of the symmetric sector.
Normally one would expect that that the perturbation of the initial data 
would lead to evolution of the system that may be considered (at least 
for small enough times) as a perturbation of GR evolution.
All contributions from the antisymmetric sector to the symmetric sector 
should appear at second-order, and the evolution equations would not 
normally drive the perturbations to become large; indeed,  this is 
precisely what the naive linearization of the system leads one to believe.

Instead one finds in general that the exact field equations give 
contributions to the symmetric sector that appear at background order 
({i.e.}, not as a small correction, but of the same order as GR 
effects), and the field equations that determine the evolution of the 
perturbations result in arbitrarily large velocities.
This not only invalidates the use of the linear approximation, but also 
explicitly shows a Cauchy instability in mNGT.

To begin, we will show how the results of the linearizations given by 
Moffat (1995a, b) and Clayton (1996a) (about a fixed GR background) 
may be retrieved by a naive (and incorrect) assumption on the Lagrange 
multiplier fields in the Hamiltonian picture, or by dropping 
acceleration terms in the antisymmetric sector field equations that 
appear at higher-order, thereby revealing a constraint that does not 
properly exist in the theory.

The relevant field configurations are those in which the antisymmetric 
components of the fundamental tensor may be considered as perturbations 
of dominant GR (symmetric sector) components on the initial data 
hypersurface $\Sigma_0$.
Thus, an expansion in powers of the antisymmetric components 
($B^a,\gamma^{[ab]}$) about the symmetric components $\gamma^{(ab)}$ 
will be made, in which the components of the fundamental tensor take 
on the form:
\begin{eqnarray}\label{eq:metric expansion}
\eqalign{
\sqrt{-{\rm g}}\sim \sqrt{S},\qquad
\beta_a\sim B_a:=S_{ab}B^b,\qquad
\alpha_a\sim -\gamma_{[ab]}B^b,\\
G_{(ab)}\sim S_{ab},\qquad
G_{[ab]}\sim -\gamma_{[ab]}:=-S_{ac}S_{bd}\gamma^{[cd]}.}
\end{eqnarray}
Terms of order $n$ in ($B^a,\gamma^{[ab]}$) will be indicated by 
${\cal O}({\rm skew}^n)$, spatial indices will be raised and lowered 
using $\gamma^{(ab)}$ and its inverse $S_{ab}$, and `$\sim$' will 
indicate the dominant contribution in powers of ($B^a,\gamma^{[ab]}$).

The assumption is that all of the antisymmetric sector tensors ($B^a$, 
$\overline{W}_a$, $\gamma^{[ab]}$, $j_{ab}$, $W$, $b_a$, $v^a_{\;b}$, 
$\lambda^a_{bc}$) are ${\cal O}({\rm skew}^{1})$, therefore leaving the 
dominant ${\cal O}(\text{skew}^0)$ terms of the symmetric sector 
Lagrange multipliers identical to those of GR:
\numparts\label{eq:LMlin}
\begin{equation}\label{eq:GR LMlin}
\fl
\Gamma\sim 0,\quad
c_a\sim 0,\quad
\sigma^a\sim \gamma^{(ab)}a_b,\quad
u^a_{\;b}\sim k^a_{\;b}\sim \gamma^{(ac)}k_{bc},\quad
\nabla^{(3)}_c[\gamma]^{(ab)}\sim 0,
\end{equation}
and those in the antisymmetric sector:
\begin{eqnarray}\label{eq:Skew LMlin}
\eqalign{
W\sim\case{4}{3}\nabla^{(3)}_a[B]^a,\qquad
v^a_{\;b}\sim\nabla^{(3)}_b[B]^a
-\case{1}{3}\delta^a_b\nabla^{(3)}_c[B]^c+j^a_{\;b},\\
\lambda^c_{ab}\sim-\case{1}{2}\gamma^{(cd)}
(\nabla^{(3)}_b[\gamma]_{[da]}+\nabla^{(3)}_a[\gamma]_{[bd]}
-\nabla^{(3)}_d[\gamma]_{[ab]})\\
\qquad +B_a{k_b}^c-B_b{k_a}^c
-\case{1}{2}\delta^c_{[a}\overline{W}_{b]}+\delta^c_{[a}b_{b]}.}
\end{eqnarray}
\endnumparts
The trace of the last of these results in the constraints
\begin{equation}\label{eq:lin bcon}
\fl
\chi_1^a\approx\case{1}{2}\gamma^{(ab)}\overline{W}_b
+\nabla^{(3)}_b[\gamma]^{[ab]}+kB^a-k^a_{\;b}B^b
\approx\case{1}{2}\gamma^{(ab)}\overline{W}_b
+\case{1}{2}{}^1\!\Xi^a\sim 0,
\end{equation}
which are equivalent to (\ref{eq:full b}) after dropping all 
contributions higher than ${\cal O}({\rm skew}^{1})$, and where the 
linear order contribution to $\Xi^a$ in (\ref{eq:Xi def}) has been 
identified as $(1/2){}^1\!\Xi^a\sim\nabla^{(3)}_b[\gamma]^{[ab]}
+kB^a-k^a_{\;b}B^b$.

Note that in dropping the contributions from the Lagrange multiplier 
$b_a$, equation (\ref{eq:full b}) has been required to play a very different 
role in the linearized system than in the nonperturbative case.
Whereas in the exact treatment, the relation (\ref{eq:full b}) is a 
condition determining the Lagrange multiplier $b_a$, in this 
linearization it has become a constraint on the Cauchy data, and $b_a$ 
left at this stage as an undetermined Lagrange multiplier.
As $\chi_1^a$ are constraints, they must be preserved in time (to this 
order) for the evolution of the system to be consistent.
It is straightforward to show that all contributions to $\dot{\chi}_1^a$ 
involving $b_a$ cancel, resulting in an additional three constraints 
$\chi^a_2:=\dot{\chi}_1^a\sim 0$.
Using the constraint (\ref{eq:lin bcon}) to determine $\overline{W}_a$, 
the linear contribution from (\ref{eq:metric evolution:b}) clearly 
depends on $b_a$, as does (\ref{eq:p dots:b}), as it takes on the form
\begin{equation}\label{eq:pdb2}
\dot{\overline{W}}_a\approx \pounds^{(3)}_{\vec{N}}[\overline{W}]_a
-2N\,{}^1\!{\cal Z}_{[a\perp]}-m^2NB_a,
\end{equation}
where the linear order contribution to (\ref{eq:Za aperp}) is 
($R^{(3)}_{ab}$ is the ${\cal O}(\text{skew}^0)$ 
GR Ricci tensor)
\begin{eqnarray}\label{eq:Za aperp1}
\eqalign{
\fl
{}^1\!{\cal Z}_{[a\perp]}\sim 
-R^{(3)}_{ab}B^b
+k_a^{\;b}k_b^{\;c}B_c-B_ak^b_{\;c}k^c_{\;b}
+k_{ab}\gamma^{[bc]}a_c-2j_{ab}a^b\\
+\case{1}{2}k^b_{\;[a}\overline{W}_{b]}-3k^b_{\;[a}b_{b]}
-(\nabla^{(3)}_a[B]^b-\delta^b_a\nabla^{(3)}_c[B]^c)a_b\\
-\gamma^{(bc)}\nabla^{(3)}_b[j]_{ac}
+\nabla^{(3)}_c[k_{ab}\gamma^{[bc]}]+k^{bc}\nabla^{(3)}_b[\gamma]_{[ca]}.}
\end{eqnarray}

Considering (\eqpdots) as second-order field equations for 
the configuration space variables, (\ref{eq:lin bcon}) determines $b_a$ as
\begin{equation}\label{eq:config b}
b_a\sim \case{1}{2}W_a-\case{1}{2}{}^1\!\Xi_a;
\end{equation} 
however, inserting this into (\ref{eq:p dots:b}) yields the algebraic 
constraint (which is in fact equivalent to $\dot{\chi}_{2a}\sim 0$)
\begin{equation}\label{eq:con const}
\dot{\overline{W}}_a\sim-{}^1\!\dot{\Xi}_a
\sim \pounds^{(3)}_{\vec{N}}[{}^1\!\Xi]_a-2N\,{}^1\!{\cal Z}_{[a\perp]}
-m^2NB_a,
\end{equation}
instead of the 
expected evolution equation.
(The field $W_a$ is easily determined in terms of $\dot{{\bf B}}^a$ 
using the linear contribution from (\ref{eq:metric evolution:b}), and 
the time derivatives on $k_{ab}$ that occur in ${}^1\!\Xi_a$ are removed 
using (\ref{eq:p dots:c}).)
The equations (\ref{eq:con const}) then play a similar role as the 
time-space components of the field equations in the massive Kalb-Ramond 
theory: $\partial_\alpha F^{0a\alpha}+m^2h^{[0a]}=0$, which do not 
involve acceleration terms and are therefore constraint equations.

One could take this system seriously and determine whether the 
constraint algebra closes properly by computing $\dot{\chi}_2^a$, 
presumably leading to a determination of $b_a$ instead of further 
constraints.
This would recover a system analogous to the massive Kalb-Ramond field, 
and, consequently, the linear approximation of Moffat (1995a, b) 
and Clayton (1996a).
A more systematic description has not been given since, as we shall see, 
this system is {\em not} an accurate representation of weak 
antisymmetric sector dynamics.
It will be shown that no additional constraints properly exist, and the 
majority of choices of arbitrarily small antisymmetric sector Cauchy 
data result in singular evolution.
Avoiding such configurations is possible by a particular choice of 
initial data, but in that case the system does not evolve in a manner 
consistent with the above linearization.

Considering the exact second-order system, (\ref{eq:metric evolution:b}) 
may be solved exactly for $W_a$ in terms of Cauchy data as
\begin{equation}\label{eq:W padot}
W_a=\frac{4}{3}\frac{1}{N\sqrt{-{\rm g}}}S_{ab}
(\dot{{\bf B}}^b-\pounds^{(3)}_{\vec{N}}[{\bf B}]^b
+Ne_c[\bgamma]^{[bc]}+N\bgamma^{[bc]}{C_{\perp c}}^\perp),
\end{equation}
which, combined with (\ref{eq:b solution}), may be used in order to 
determine the exact form of the acceleration terms appearing in 
(\ref{eq:p dots:b}) to be 
\begin{equation}\label{eq:W acc}
\dot{\overline{W}}_a\sim 
\frac{4}{3}\frac{1}{N\sqrt{-{\rm g}}}O^{-1}_{1ab}O_2^{bc}S_{cd}
\ddot{{\bf B}}^d;
\end{equation}
the time derivatives on $k_{ab}$ and $j_{ab}$ that appear are once again 
removed by making use of (\ref{eq:p dots:c}). 
Clearly to lowest order $O^{ab}_1\sim \gamma^{(ab)}$, and so its inverse 
is $O^{-1}_{1ab}\sim S_{ab}$.
The operator $O_2^{ab}$ in (\ref{eq:Os}) on the other hand, is 
${\cal O}({\rm skew}^2)$
\numparts
\begin{equation}\label{eq:O}
O_2^{ab}\sim B^aB^b
+\case{1}{4}\gamma^{[ac]}\gamma_{[cd]}\gamma^{(db)}
= B^aB^b-\gamma^{(ab)}\gamma\cdot\gamma+\gamma^a\gamma^b,
\end{equation}
where (\ref{eq:metric expansion}) has been used in order to deduce that 
$G_a\sim-\gamma_a$ in (\ref{eq:exact defns}), also leading to the form 
of the inverse from (\ref{eq:left inv}) 
\begin{equation}\label{eq:inverse O2}
O^{-1}_{2ab} \sim -\frac{1}{\gamma\cdot\gamma}
\Bigl[\gamma_{(ab)}
-\frac{1}{\gamma\cdot B}(B_a\gamma_b+\gamma_aB_b)
+\frac{B\cdot B-\gamma\cdot\gamma}
{(\gamma\cdot B)^2}\gamma_a\gamma_b\Bigr],
\end{equation}
\endnumparts
which is clearly ${\cal O}({\rm skew}^{-2})$.
The presence of the operator $O^{ab}_2$ in (\ref{eq:W acc}) immediately 
shows how these acceleration terms disappear in the above linearized 
analysis, since it causes the acceleration terms in (\ref{eq:W acc}) 
to appear at third-order, leaving these field equations as constraints 
which do not truly exist in the theory.

Furthermore (\ref{eq:W acc}) will result in very poorly behaved 
accelerations near configurations where the antisymmetric sector Cauchy 
data is small unless an {\em ad hoc} condition is imposed on the initial 
data, leaving a nontrivial wave equation for ${\bf B}^a$.
To see this, (\ref{eq:p dots:c}) may be written to 
${\cal O}(\text{skew}^3)$ as:
\begin{equation}\label{eq:pdc}
\frac{4}{3}\frac{1}{N\sqrt{S}}O_{2ab}\ddot{{\bf B}}^b+{}^1\!\dot{\Xi}_a
\sim\pounds^{(3)}_{\vec{N}}[{}^1\!\Xi]_a
-2N\,{}^1\!{\cal Z}_{[a\perp]}-Nm^2B_a
+{}^3\Psi_a,
\end{equation}
where ${}^3\Psi_a$ are the remaining non-acceleration contributions to 
the field equations.
The acceleration of ${\bf B}^a$ is therefore given by
\begin{equation}\label{eq:B acc}
\ddot{{\bf B}}^a
\sim \case{3}{4}N\sqrt{S}{O_2^{-1}}^{ab}(-2\chi_{2b}+{}^3\Psi_b),
\end{equation}
where $\chi_{2a}$ has been identified from the time evolution of 
(\ref{eq:lin bcon}).
Due to the presence of ${O_2^{-1}}^{ab}$ in (\ref{eq:B acc}), the right 
hand side is generally ${\cal O}(\text{skew}^{-1})$, resulting in 
unboundedly large accelerations resulting from arbitrarily small 
antisymmetric sector Cauchy data on $\Sigma_0$.
(Note that the sign of the right hand side of (\ref{eq:B acc}) is 
uncorrelated with that of $B^a$ in general, since the sign of $O_2^{ab}$ 
may be chosen by adjusting the relative magnitudes of $B^a$ and 
$\gamma^{[ab]}$ on $\Sigma_0$.) 
To avoid this, one could require that the initial data satisfy a 
condition that mimics the linearized result (\ref{eq:lin bcon})
\begin{equation}\label{eq:CD cond}
\chi_{2a}\sim{}^3\!\psi_a,
\end{equation}
effectively constraining the linear order contributions to 
(\ref{eq:B acc}) to vanish up to terms of higher-order.
However even in this case, (\ref{eq:B acc}) becomes
\begin{equation}\label{eq:B acc:corr}
\ddot{{\bf B}}^a
\sim\case{3}{4}N\sqrt{S}{O_2^{-1}}^{ab}({}^3\!\psi_b+{}^3\Psi_b),
\end{equation}
which is a nontrivial ${\cal O}(\text{skew}^1)$ evolution equation for 
${\bf B}^a$ that is not reproduced by the naive linearization.
(Note that the second order spatial derivatives that 
combine with the time derivatives of ${\bf B}^a$ to make 
(\ref{eq:p dots:b}) a hyperbolic wave equation have also dropped out of 
${\cal Z}_{[a\perp]}$ at linear order in (\ref{eq:Za aperp1}), 
presumably reappearing at third order to give (\ref{eq:B acc:corr}) 
the appropriate hyperbolic form.)
  
From the first-order point of view, the relation (\ref{eq:b sol}) 
determines the Lagrange multiplier $b_a$ in all cases when the operator 
$O^{ab}_2$ is nondegenerate, including weak antisymmetric sector field 
configurations. 
(If $O^{ab}_2$ is degenerate then it may only be inverted on some 
subspace, determining some components of $b_a$ and leaving the rest 
undetermined.
The analysis of these cases would be rather more complicated and will 
not be discussed further.)
Since, as can be readily seen from (\ref{eq:left inv}), $O^{-1}_{2ab}$ 
is ${\cal O}(\text{skew}^{-2})$ and therefore generally becomes singular 
for vanishingly small antisymmetric sector, the presence of $b_a$ in 
(\ref{eq:metric evolution:b}) would drive ${\bf B}^a$ to evolve 
arbitrarily quickly as one considers vanishingly small antisymmetric sector 
Cauchy data.
Explicitly, the last term in (\ref{eq:metric evolution:b}) may be 
replaced using (\ref{eq:b sol}) leaving
\begin{equation}\label{eq:B ev:corr}
\fl
\dot{{\bf B}}^a\approx\pounds^{(3)}_{\vec{N}}[{\bf B}]^a
-N(\nabla^{(3)}_b[\bgamma]^{[ab]}+\bgamma^{[ab]}{C_{\perp b}}^\perp)
+\case{3}{4}N\bgamma^{(ab)}O^{-1}_{2bc}(O_1^{cd}\overline{W}_d+\Xi^c).
\end{equation}
The presence of $O^{-1}_{2ab}$ in the last term indicates that 
unboundedly large velocities generally occur for infinitesimally small 
antisymmetric sector Cauchy data on $\Sigma_0$, and it is easy to see 
that the presence of $b_a$ in ${\cal Z}_{[a\perp]}$ will lead to similar 
behaviour for (\ref{eq:p dots:b}).
Once again choosing initial data such that (\ref{eq:CD cond}) is 
satisfied removes all of these effects, as then (\ref{eq:b sol}) 
becomes (to ${\cal O}(\text{skew}^3)$)
\begin{equation}\label{eq:bsol:corr}
b_a\sim\case{1}{2}\overline{W}_a
+\case{1}{2}O^{-1}_{2ab}(2\,{}^3\!\psi^b-B^bB^c\overline{W}_c
+{}^3\!\Xi^b),
\end{equation}
effectively resulting in $b_a\sim{\cal O}(\text{skew}^3)$.
(Note that this alone does not guarantee that $b_a$ vanishes smoothly 
with the vanishing of the antisymmetric sector, but it is 
straightforward to choose data such that it does; for example, one could 
choose $O_1^{ab}\overline{W}_b+\Xi^a\approx O_2^{ab}f_b$, where $f_b$ 
is an ${\cal O}(\text{skew})$ covector field.

Although we have been focusing on the dynamics of 
$(B^a,\overline{W}_a)$, it is fairly straightforward to see that unless 
(\ref{eq:CD cond}) is imposed on $\Sigma_0$ guaranteeing that 
$b_a\sim{\cal O}(\text{skew}^1)$, the evolution of the symmetric sector 
will generally not resemble that of GR.
The contributions from $b_a$ to the Hamiltonian constraint and the 
symmetric sector evolution equations imply that the symmetric sector 
cannot in general be considered as perturbative corrections to GR 
dynamics, since there are contributions from the antisymmetric sector 
that show up at ${\cal O}(\text{skew}^0)$.
Even if one imposed (\ref{eq:CD cond}) on $\Sigma_0$, one would have to 
check that it is preserved in time in order for the system to remain 
well-behaved in evolution.
This calculation would depend strongly on the chosen form of 
$\,{}^3\!\psi$ and will not be pursued here.
Nevertheless, it is clear that if one makes the choice 
(\ref{eq:CD cond}), the resulting dynamics {\em cannot} be well 
approximated by the linearized system, simply because there are no 
additional constraints appearing.
The evolution equation (\ref{eq:metric evolution:b}) is {\em not} a 
constraint, and neither is (\ref{eq:p dots:b}), showing that the full 
six degrees of freedom in the antisymmetric sector propagate even in the 
weak field regime, contradicting the linearized results.
This behaviour in the antisymmetric sector indicates that the linearization 
given by Moffat (1995a, b) and Clayton (1996a) does not represent 
the weak-field evolution as determined from the full field equations, 
and therefore cannot be trusted: mNGT is not linearization stable about 
GR backgrounds.

It is clear that these instabilities occur whenever the Lagrange multiplier 
$b_a$ becomes singular as the perturbations of the Cauchy data vanish.
The general form of the solution of (\ref{eq:full b}) specialized to 
spherically symmetric systems is
\begin{equation}\label{eq:SS b}
b_1\approx \frac{1}{2}\frac{\gamma^{11}\overline{W}_1}{(B^1)^2}
+2\frac{k^2_{\;2}}{B^1},
\end{equation} 
and one finds that in regions of spacetime where $B^1$ is vanishingly 
small (for example, in an exact GR background or in the asymptotic 
region of an arbitrary asymptotically flat spacetime), perturbations 
generally cause $b_1$ to become arbitrarily large, which in turn cause 
antisymmetric sector velocities to become large.
Thus although the Wyman sector solution (Wyman 1950; Cornish 1994) (which is becoming the basis for much of the phenomenology of 
NGT (Moffat and Sokolov 1995a, b)) assumes that both 
$B^1$ and $\overline{W}_1$ vanish globally, if one considers 
perturbations of these fields on $\Sigma_0$, the above behaviour 
reappears.
Therefore, although the instabilities have been proven to exist for 
configurations very close to purely GR spacetimes, one expects that 
any asymptotically flat spacetime will also suffer them.
In particular, since there is no asymptotically flat spacetime in 
which $(B^1,\overline{W}_1)$ is asymptotically nonvanishing (Clayton 
1996a), the unique asymptotically flat, static spherically symmetric 
solution with nontrivial antisymmetric sector (the Wyman solution) will 
also be unstable against perturbations.

This is essentially the same effect as was found in (Isenberg and 
Nester 1977), where the effect of gravitational dynamics on the 
constraints of various derivative coupled vector fields was studied.
It was found that constraints on the vector field may be lost when GR 
is considered as evolving concurrently with the vector field (as 
opposed to the vector field evolving on a GR background).
This manifested itself as an increase in the number of degrees of 
freedom in the vector field, and singular behaviour in the evolution 
equations when approaching asymptotically flat spacetimes.
There one finds no evidence of this when considering the vector field 
dynamics on a fixed GR background, which is analogous to the linear 
approximation here.

Kucha\v{r} (1977) has noted that fields such as these that are derivatively 
coupled to GR (as NGT may be considered to be) may propagate off of the light 
cone as determined by the spacetime metric, and indeed, there are explicit 
examples of such behaviour (Buchdahl 1958, 1962; Cohen 1967; Velo and 
Zwanzinger 1969; Aragone and Deser 1971).
It is therefore not surprising that the field equations of UFT (and hence 
NGT) have been found to propagate information in a manner that is inconsistent 
with a simple causal structure as determined from a single spacetime metric; 
in fact Maurer-Tison (1959) has found that in UFT there are three such metrics
that one must take into account.
In UFT (or massless NGT), these metrics were consistent in the sense that 
there was locally a largest light cone (although the metric that defined this 
would change depending on the strength of the antisymmetric sector), and in 
the limit of vanishing antisymmetric sector all three degenerated to the 
single causal metric of GR (Maurer-Tison 1956).
Although these metrics are not known for mNGT at this time, given the fact 
that the acceleration terms for ${\bf B}^a$ have no simple GR limit, it is 
reasonable to suspect that one of these light cones has been altered so as to 
destroy the above mentioned consistency (perhaps resulting in a light cone 
that is degenerate in the GR limit).

\section{Conclusions\label{sect:concl}}

It has been shown that arbitrarily small antisymmetric sector Cauchy 
data leads to singular evolution for the majority of possible choices of 
perturbatively small antisymmetric sector initial data.
The results followed from an examination of the exact field equations 
of the massive nonsymmetric gravitational theory near purely general 
relativistic field configurations, and has been demonstrated by 
considering both the first and second order points of view, showing 
how the naive linearization fails to accurately describe the system, 
even in those cases where the choice of initial data does not lead to 
singular evolution.
In doing so, it has been shown the constraints that guaranteed good 
fall-off for the linearized fields do not properly exist even for weak 
fields.
Given this, it seems that the criterion for choosing the form of the 
mNGT action (Clayton 1996a) has not truly been fulfilled, and it 
remains unclear whether one has truly made an improvement over the 
original formulation of NGT.

The failure of linearization stability has been noted in GR for closed 
spaces (Brill and Deser 1973).
Here it has been found that not only can the linearized system not be 
trusted, but also that what appear to be benign perturbations of 
particular initial data ({i.e.}, as having very little effect on the 
evolution of the system as a whole) result in very different evolution, 
in which some of the antisymmetric sector fields are given arbitrarily 
large velocities.
This is clearly interpretable as a Cauchy instability in the usual 
sense, since the evolution of generic configurations that are 
arbitrarily `close' to a GR spacetime (or part thereof) does not 
smoothly depend on the initial data, and makes it difficult to 
physically interpret such spacetimes.
Denying the physical importance of these configurations amounts to 
labelling as unphysical purely GR spacetimes and asymptotically flat 
spacetimes, and a Newtonian limit would instead have to be recovered 
in some (presumably stable) region of spacetime in which the 
antisymmetric sector is not small compared to the symmetric sector.
Clearly if one does not allow asymptotically flat (or nearly so) 
spacetimes, nor regions of spacetime where the symmetric sector 
dominates to the point where one essentially recovers GR physics 
locally, then these instabilities are avoided.

Despite the fact that these results indicate that the relevant (Wyman) 
solution would be unstable, there are some encouraging phenomenological 
results on galaxy dynamics (Moffat and Sokolov 1995a), as well as some 
optimism that the collapse of spherically symmetric matter would be 
nonsingular (Moffat and Sokolov 1995b).
It is possible to further modify the dynamics of NGT in order to remove 
these instabilities and thereby recover this phenomenology by 
guaranteeing that three of the field equations appear as constraints 
rigorously, either imposed via Lagrange multipliers in the action 
(Moffat 1996), or leading to generalizations of the model introduced 
by Damour \etal (1993) that possess `gauge invariant kinetic terms'.
Alternatively, given that the antisymmetric sector should appear as a 
Kalb-Ramond field (Kalb and Ramond 1974), it is reasonable to expect 
that there should be some tie between NGT and string theory 
(Moffat 1995d).
It may then be the case that the fundamentally nonlocal nature of 
strings (which does not have a Cauchy initial value formulation) would 
cause the NGT field equations to be an inadequate description of the 
system in some situations.

\ack
I would like to thank my supervisor J. W. Moffat and the University of 
Toronto for supporting this research, J. L\'{e}gar\'{e} for his careful 
reading this manuscript, and J. W. Moffat and L. Demopoulos for 
comments and discussions related to this work.

\References
\item 
Aragone C and Deser S 1971 Il Nuovo Cimento {\bf 3A} 709--20

\item
Arnowitt R, Deser S and Misner C W 1959 Phys. Rev. {\bf 116} 1322--30

\item 
Bonnor W B 1954 Proc. R. Soc. London {\bf A226} 366--77

\item
Brill D and Deser S 1973 Comm. Math. Phys. {\bf 32} 291--304

\item 
Buchdahl H A 1958 Il Nuovo Cimento {\bf 10} 96--103

\item
\dash 1962 Il Nuovo Cimento {\bf 25} 486--96

\item
Callaway J 1953 Phys. Rev. {\bf 92} 1567--70

\item
Clayton M A 1995 {\it Hamiltonian formulation of nonsymmetric 
gravitational theories} University of Toronto Preprint UTPT-95-20

\item
\dash 1996a J. Math. Phys. {\bf 37} 395--420

\item
\dash 1996b {\it Massive Nonsymmetric Gravitational Theory: {A} 
{H}amiltonian Approach} Ph.D. thesis, Department of Physics, University 
of Toronto

\item
Cohen H A 1967 Il Nuovo Cimento {\bf 52A} 1242--52

\item
Cornish N J 1994 Mod. Phys. Lett. A {\bf 9} 3629--40

\item
Damour T, Deser S and McCarthy J 1992 Phys. Rev. D {\bf 45} R3289--91

\item
\dash 1993 Phys. Rev. D {\bf 47} 1541--56

\item
Einstein A 1945 Ann. Math. {\bf 46} 578--84

\item
Einstein A and Straus E G 1946 Ann. Math. {\bf 47} 731--41

\item
Hawking S W and Ellis G F R 1973 {\it The Large Scale Structure of 
Space-Time} (Cambridge: Cambridge Monographs on Mathematical Physics, 
Cambridge University Press)

\item
Infeld L 1950 Acta. Phys. Pol. {\bf 10} 284--93

\item
Isenberg J A and Nestor J M 1977 Ann. Phys. {\bf 107} 56--81

\item
\dash 1980 {\it General Relativity and Gravitation: One Hundred Years 
After the Birth of Albert Einstein} vol~I ed Held A (New York: Plenum 
Press) p 23--97

\item
Kalb M and Ramond P 1974 Phys. Rev. D {\bf 9} 2273--84

\item
Kelly P F 1991 Class. Quantum Grav. {\bf 8} 1217--29

\item
\dash 1992 Class. Quantum Grav. {\bf 9} 1423, Erratum

\item Kucha\v{r} K 1977 J. Math. Phys. {\bf 18} 1589--97

\item
Kunstatter G, Leivo H P and Savaria P 1984 Class. Quantum Grav. {\bf 1} 
7--13

\item
L\'{e}gar\'{e} J and Moffat J W 1995 Gen. Rel. Grav. {\bf 27} 761--75

\item
Maurer-Tison F 1956 C. R. Acad. Sc. {\bf 242} 1127--9

\item
\dash 1959 Ann. Scient. \'{E}c. Norm. Sup. {\bf 76} 185--269

\item
Moffat J W 1979 Phys. Rev. D {\bf 19} 3554--58

\item
\dash 1984 Found. Phys. {\bf 14} 1217--52

\item
\dash 1991 {\it Gravitation: A Banff Summer Institute} ed Mann R and 
Wesson P (New Jersey: World Scientific) p 523-97

\item
\dash 1995a Phys. Lett. B {\bf 355} 447--52

\item
\dash 1995b J. Math. Phys. {\bf 36} 3722--32

\item
\dash 1995c J. Math. Phys. {\bf 36} 5897--915

\item
\dash 1995d {\it Nonsymmetric gravitational theory as a string theory} 
University of Toronto Preprint UTPT-95-26

\item
\dash 1996 {\it Dynamical constraints in the nonsymmetric gravitational
  theory} University of Toronto Preprint UTPT-96-05

\item
Moffat J W and Sokolov I Yu 1995a {\it Galaxy dynamics in the nonsymmetric
  gravitational theory} To appear in Phys. Lett. B

\item
\dash 1995b {\it On gravitational collapse in the nonsymmetric 
gravitational theory} University of Toronto Preprint UTPT-95-21

\item
Teitelboim C 1973 Ann. Phys. {\bf 79} 542--57

\item
\dash 1980 {\it General Relativity and Gravitation: One Hundred Years 
After the Birth of Albert Einstein} vol~I ed Held A (New York: Plenum 
Press) p 195--225  

%
\item Velo G and Zwanzinger D 1969 Phys. Rev. {\bf 188} 2218-22

\item
Wyman M 1950 Can. J. Math. {\bf 2} 427--39
\endrefs

\end{document}